\documentclass[aps,prd,superscriptaddress,floatfix,nofootinbib,preprintnumbers,eqsecnum,twocolumn]{revtex4-2}

\pdfoutput=1
\usepackage{amssymb,amsfonts,amsmath,graphicx,bbm}
\usepackage{comment}
\usepackage{xcolor}
\usepackage{orcidlink}
\usepackage[english]{babel}
\usepackage[T1]{fontenc}
\usepackage{amsmath}
\usepackage{slashed}
\usepackage{booktabs}
\usepackage{listings}
\usepackage[utf8]{inputenc}
\usepackage{circuitikz}
\usepackage{nicematrix}
\usepackage{braket}
 \hypersetup{ 
   colorlinks,
   linkcolor={blue!80!black},
   citecolor={blue!70!black},
   urlcolor={blue!70!black}
 }
\usepackage{mathtools}
\usepackage{footmisc}
\usepackage{mathrsfs}
 
\usepackage{placeins}

\usepackage{amssymb,amsmath,epsfig,longtable}

\numberwithin{equation}{section}

\newcommand{\bean}{\begin{eqnarray*}}
\newcommand{\eean}{\end{eqnarray*}}

\newcommand{\cN}{{\cal N}}

\newcommand{\cA}{{\cal A}}
\newcommand{\cB}{{\cal B}}
\newcommand{\cC}{{\cal C}}

\newcommand{\cV}{{\cal V}}

\def\cjn1{{\cA, \cC^*\otimes \wedge^j \cN^*}}
\def\bjn1{{\cA, \cB^*\otimes \wedge^j \cN^*}}
\def\vjn1{{\cA, \cV^*\otimes \wedge^j \cN^*}}
\def\cjn2{{\cA, \cC\otimes \wedge^j \cN^*}}
\def\bjn2{{\cA, \cB\otimes \wedge^j \cN^*}}
\def\vjn2{{\cA, \cV\otimes \wedge^j \cN^*}}

\newcommand{\be}{\begin{equation}}
\newcommand{\ee}{\end{equation}}
\newcommand*{\nnbe}{\begin{equation}}
\newcommand*{\nnee}{\end{equation}}
\newcommand{\bea}{\begin{eqnarray}}
\newcommand{\eea}{\end{eqnarray}}
\newcommand{\ba}{\begin{align}}
\newcommand{\ea}{\end{align}}
\newcommand{\bi}{\begin{itemize}}
\newcommand{\ei}{\end{itemize}}

\newsavebox{\overlongequation}

\renewcommand{\arraystretch}{1.7}

\usepackage{hyperref}

\begin{document}
\title{Not So Flat Metrics}


\author{Kit Fraser-Taliente}
\email[]{cristofero.fraser-taliente@physics.ox.ac.uk}
\affiliation{Rudolf Peierls Centre for Theoretical Physics, University of Oxford, Parks Road, Oxford OX1 3PU, UK}

\author{Thomas R. Harvey}
\email[]{trharvey@mit.edu}
\affiliation{The NSF AI Institute for Artificial Intelligence and Fundamental Interactions, Massachusetts Institute of Technology, 77 Massachusetts Avenue, Cambridge, MA 02139, USA}
\affiliation{Center for Theoretical Physics, Massachusetts Institute of Technology, 77 Massachusetts Avenue, Cambridge, MA 02139, USA
}

\author{Manki Kim}
\email[]{mk2427@stanford.edu}
\affiliation{Stanford Institute for Theoretical Physics, 382 Via Pueblo, Stanford, CA 94305, USA}

%\date{\today}

\begin{abstract}
In order to be in control of the $\alpha'$ derivative expansion, geometric string compactifications are understood in the context of a large volume approximation. In this letter, we consider the reduction of these higher derivative terms, and propose an improved estimate on the large volume approximation using numerical Calabi-Yau metrics obtained via machine learning methods. Further to this, we consider the $\alpha'^3$ corrections to numerical Calabi-Yau metrics in the context of IIB string theory. This correction represents one of several important contributions for realistic string compactifications -- alongside, for example, the backreaction of fluxes and local sources -- all of which have important consequences for string phenomenology. As a simple application of the corrected metric, we compute the change to the spectrum of the scalar Laplacian.

\end{abstract}

%\keywords{}
\maketitle		
%\section{}
%\subsection{}
%\subsubsection{}

%%%%%%%%%%%%%%%%%%%%%%%%%%%%%%%%%%%%%%%%%%%%%%%%%%%%%%%%%%%%%%%%%%%%%%%%%%
%%%%%%%%%%%%%%%%%%%%%%%%%%%%%%%%%%%%%%%%%%%%%%%%%%%%%%%%%%%%%%%%%%%%%%%%%%

\section{Introduction}
In recent years, there has been substantial work on computing numerical Ricci-flat metrics on Calabi-Yau (CY) manifolds~\cite{donaldson2005numerical, Braun:2007sn, Douglas:2006rr, Headrick:2009jz}, primarily using machine learning techniques~\cite{Ashmore:2019wzb,Anderson:2020hux, Jejjala:2020wcc, Douglas:2020hpv,Larfors:2021pbb,Larfors:2022nep,Ashmore:2021ohf,Gerdes:2022nzr}. These have been used to calculate previously inaccessible quantities of the resulting low energy theories~\cite{Braun:2008jp, Ahmed:2023cnw, Constantin:2024yxh}. However, in a realistic string compactification, these internal metrics are corrected by multiple effects — in particular, the back-reaction of fluxes and local sources, alongside higher derivative $\alpha'$ corrections. Computation of all of these corrections has important consequences for the phenomenology of string compactifications. Further to this, geometric string compactifications are usually understood in the context of the large volume approximation, where the volume of the CY $V \gg (2\pi)^6 \alpha'^3$, such that one can argue that the effects of the higher derivative terms are small. Now that numerical metrics are easily available, it makes sense to impose stronger pointwise conditions on string compactifications.

Unfortunately, as emphasised by \cite{Dine:1985he}, parametric control over higher derivative corrections is not possible for compactifications with moduli stabilisation. One therefore needs either to embrace higher derivative corrections or understand precisely when such corrections can be safely ignored in the low energy theory. As we have access to the metric, a refined numerical check can be performed pointwise.

The Calabi-Yau metric and the associated $\alpha'$ corrections are important defining data of Calabi-Yau CFTs. Although some aspects of Calabi-Yau CFTs are well understood, partly due to their enhanced worldsheet supersymmetry \cite{Lerche:1989uy, Odake:1988bh,Odake:1989dm,Eguchi:1988vra,Lin:2015wcg,Lin:2016gcl,Gepner:1987qi,Gepner:1987vz}, access to non-supersymmetric physical observables is highly limited. To investigate non-supersymmetric stringy phenomena, e.g., four-point graviton amplitudes, physical Yukawa couplings~\cite{Constantin:2024yxh,Butbaia:2024tje}, the understanding of the Calabi-Yau metric appears to be inevitable.

In this letter, we begin by considering reduction of higher derivative terms, and present a general method for checking the validity of effective field theories originating from string theory. These checks are a pointwise improvement on the usual large volume approximation. We then consider a numerical calculation of one particular correction to CY metrics for type-IIB string theory, originating from the leading $\alpha'$ correction to the 10D effective action. This particular term appears at leading order in the string coupling $g_{st}=e^{\langle\phi\rangle}$~\cite{Becker:2002nn,Becker:2015bra}. As a simple application of this corrected metric, we calculate the shift to the scalar Laplacian.

We briefly discuss the choice of units throughout this text. We normalise the worldsheet action as
\begin{equation}
    S=\frac{1}{2\pi\alpha'} \int_\Sigma J\,,
\end{equation}
where $J$ is the K\"ahler form defined on the Calabi-Yau manifold, and $\alpha'$ is related to the string length via
\begin{equation}
    l_s=2\pi\sqrt{\alpha'}\,.
\end{equation}
We then normalise the dimensionless Calabi-Yau volume measured in string units to
\begin{equation}
    \mathcal{V} = \frac{V}{l_s^6} = \frac{1}{l_s^6} \int_{CY} \sqrt{g}=  \frac{1}{6l_s^6}\int J\wedge J \wedge J.
\end{equation}

In these units, the large volume condition, $V \gg l_s^6$, is rewritten in terms of the dimensionless volume as
\begin{equation}\label{eqn:largeVol}
    \mathcal V \gg 1.
\end{equation}

For the purposes of numerical geometry, it is useful to work in units such that the dimensionless volume $\mathcal V$ is 
\begin{equation}
    \mathcal{V} = 1\,.
\end{equation}
We then analytically keep track of how the various quantities scale with $\mathcal{V}$, allowing us to extrapolate the result to arbitrary volume. One should therefore not be concerned that this is incompatible with equation~\ref{eqn:largeVol}.

As we will be working with the Fermat quintic throughout, this is equivalent to setting the single K\"ahler modulus $t$ to $(6/5)^{1/3}$, where $t$ is given by
\begin{equation}
    t:= \frac{1}{l_s^2} \int_{\mathcal{C}} J\,,
\end{equation}
and $\mathcal{C}\equiv \Bbb{P}^1$ is the Hilbert basis element of the Mori cone of the quintic. Again, this is not a concern, as we can always rescale $t$ to a physically viable value.

\section{Numerically checking validity of the $\alpha'$ expansion}\label{sec:Valid}
\begin{figure}
    \centering
    \includegraphics[width=\linewidth]{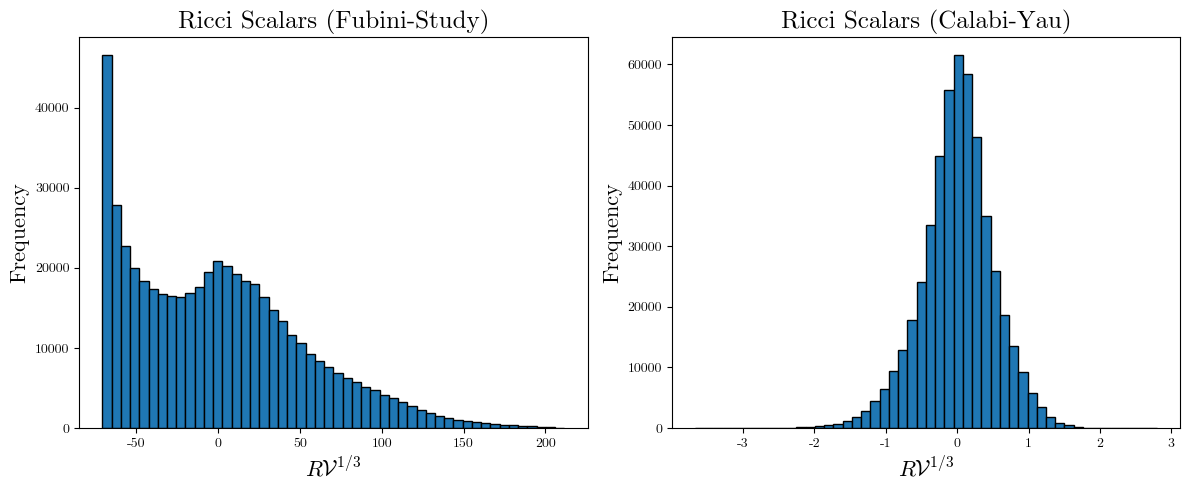}
    \caption{Histogram of $R\mathcal V^{1/3}$, in string units, where $R$ is the Ricci scalar. The plot on the left is before training the numerical metric, while the one on the right is after training.}
    \label{fig:TrainingFlat}
\end{figure}
As mentioned in the previous section, geometric string compactifications are usually understood in the context of an $\alpha'$ derivative expansion of the 10 dimensional effective field theory. Schematically, the gravity sector Lagrangian from a string theory, where we have set the dilaton to a constant, is of the form
\begin{equation}
    \mathcal{L} \propto R +  \sum_{n=L}\alpha'^{n} f_n(R_{MNPQ}),
\end{equation}
where each $f_n$ is a sum over different contractions of $n+1$ copies of the Riemann tensor $R_{MNPQ}$. The exact expressions of these higher order terms, and the order of the leading correction $L$, depend on the particular choice of string theory.

Typically, in a geometric string compactification, one truncates to leading order in $\alpha'$, implicitly assuming that the higher order corrections are negligible. CY manifolds satisfy the resulting equations, but despite being Ricci-flat, their Riemann tensor does not vanish, so there is a risk that the higher order corrections could be large. Without access to the metric on the CY manifold, one usually checks the validity of this approximation in the low energy effective theory by confirming that the dimensionful volume of the compact space $V$ satisfies $V\gg(2\pi)^6\alpha'^3$. This is  a rather crude check, as it assumes that the curvature is roughly constant across the compact space. Without access to the metric, it is difficult to conceive of any better approach.

Now that numerical Ricci-flat metrics on CY manifolds are available, we can improve on this analysis by considering pointwise checks. If one truncates the theory to $\alpha'^p$ we need only evaluate $f_n$ to order $\alpha'^{p-2n}$. As a result, the simplest pointwise condition is
\begin{equation}
    \alpha'^{L} V^{2/6} |f_L(R^0_{MNPQ})| \ll 1 ,
\end{equation}
or equivalently
\begin{equation}\label{eqn:NewCond}
    (2\pi)^2\alpha'^{L+1} \mathcal{V}^{2/6} |f_L(R^0_{MNPQ})| \ll 1
\end{equation}
where $R^0_{MNPQ}$ is the Riemann tensor calculated from the Ricci-flat metric, and the volume of the CY is included to keep the expression dimensionless.

We begin by training a depth-four, width 64 projective neural network, using \texttt{cymetric}'s $\phi$-model with $10^6$ points on the Fermat quintic. We achieve a Monge-Ampère loss of $8\times 10^{-4}$ in a matter of minutes, and the trained metric gives a Ricci scalar plotted in figure~\ref{fig:TrainingFlat}. It is encouraging to note that after training, the Ricci scalars are distributed nearly normally around zero, with $\sigma = 0.5$.

\begin{figure}
    \centering
    \includegraphics[width=\linewidth]{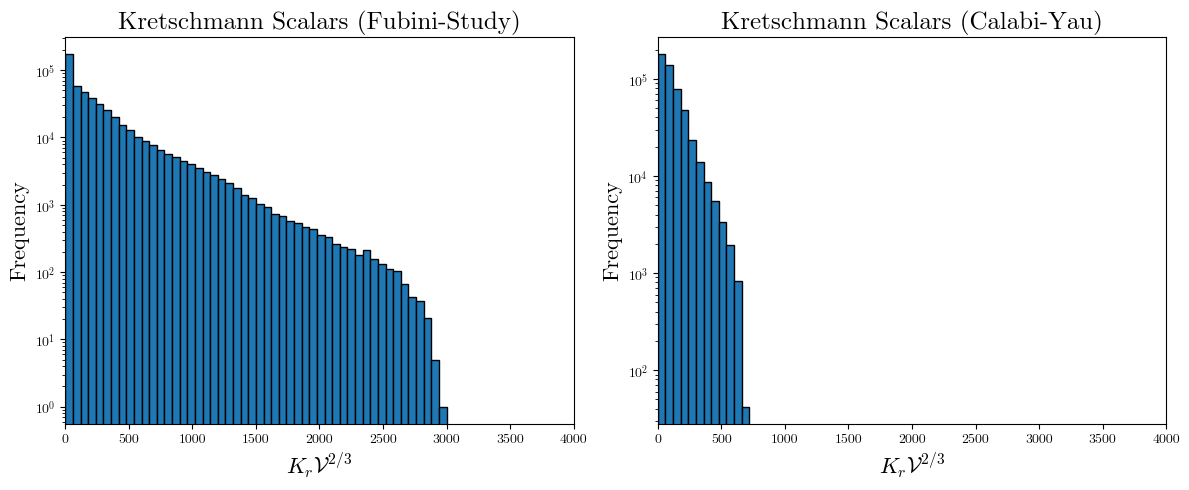}
    \caption{Histogram of $K_r \mathcal{V}^{2/3}$ across the Calabi-Yau manifold, in string units. The left plot is before training the numerical metric, and the right plot is after training.}
    \label{fig:histRSquare}
\end{figure}

\begin{figure}
    \centering
    \includegraphics[width=0.88\linewidth]{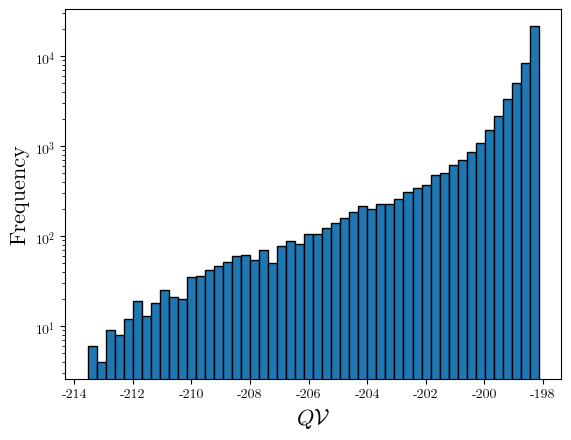}
    \caption{Histogram of $Q\mathcal{V}$ across the Calabi-Yau manifold with the numerical metric, in string units. }
    \label{fig:histQ}
\end{figure}

Rather than taking a given string model, we consider some generic higher derivative corrections. In particular, we consider the Kretschmann scalar,
\begin{equation}\label{}
    K_r = \frac{1}{2\pi}R^{\bar{m}n\bar{p}q}R_{n\bar{m}q \bar{p}},
\end{equation}
and
\begin{equation}\label{eqn:Q}
Q= \frac{1}{3(2 \pi)^3}\left({R}_{a \bar{b}}{ }^{c \bar{d}} {R}_{c \bar{d}}{ }^{e \bar{f}} {R}_{e \bar{f}}{ }^{a \bar{b}}-{R}_a{ }^c{ }_b{ }^d {R}_c{ }^e{ }_d{ }^f {R}_e{ }^a{ }_f{ }^b\right).
\end{equation}
$K_r$ occurs as the leading order correction to the Heterotic string, while $Q$, as will be discussed in the next section, is of importance to IIB string theory. The numerical factors have been chosen to match with those that appear in the respective string theories. The histograms for $K_r$ and $Q$ across the Fermat quintic CY are in figures~\ref{fig:histRSquare} and~\ref{fig:histQ}, in string units. 

Without a given string model, we cannot definitively give $\mathcal V$ a value. However, it is clear that $K_r$ and $Q$ vary across the manifold, even for a manifold as simple as the Fermat quintic. The regions where these are large may violate equation~\ref{eqn:NewCond}, despite satisfying the large volume condition. In principle, these are important checks to carry out for any geometric model constructed from string theory to confirm the validity of the low energy theory.


\section{Computing the Shift to the Metric}\label{sec:MetricShift}
Another important property of $\alpha'$ corrections is the induced modification to the Ricci-flat metric. To explore this, we now turn to the specific $\alpha'$ corrections that occur in IIB string theory, and calculate the resulting change to the metric~\cite{Becker:2002nn,Becker:2015bra}. It turns out that this correction to the metric is very amenable to being calculated with machine learning methods.

From the worldsheet perspective, to leading order in $\alpha'$, the metric $\tilde{G}$ of the compact space $X$ satisfies (in complex coordinates $z^m,\overline{z}^{\overline m}$):
\begin{align}
    \beta^G_{mn} &= R_{mn} = 0, \\ \beta^G_{\overline{mn}} &= R_{\overline{mn}} = 0, \\ \beta^G_{m\overline{n}} &= R_{m\overline{n}} = \frac{1}{2\pi} \partial_{m}\partial_{\overline{n}}{\rm Tr}(\log \tilde{G})=0.
\end{align}
As a result $\tilde{G}$ is Kähler and Ricci flat,

Continuing to the next order in $\alpha'$, the purely holomorphic and anti-holomorphic beta functions remain zero, but the mixed terms are corrected by
\begin{equation}\label{eq:Corrected}
    \frac{1}{2\pi}\partial_m\partial_{\overline n}(\log(G)) + \frac{(2\pi \alpha')^3}{8\pi} \zeta(3)\partial_m\partial_{\overline n} Q(G) = 0,
\end{equation}
where $G$ is the corrected metric, and $Q$ is the 6D Euler density for Ricci-flat manifolds 
\begin{equation}
\int_{X}\text{dVol}_{\tilde{G}}Q({\tilde{G}}) = \chi. 
\end{equation}
For Kähler manifolds, this is given by equation \ref{eqn:Q}.
To proceed, we make the following ansatz:
\begin{equation}
    \frac{1}{2\pi}(\log(G)) + \frac{(2\pi \alpha')^3}{8\pi} \zeta(3) Q(G) = \frac{1}{2\pi}(\log(\tilde G)) + f(z) + g(\overline{z}),
\end{equation}
where $f$ and $g$ are arbitrary holomorphic and anti-holomorphic functions respectively. Clearly, if true, the above equation implies~\eqref{eq:Corrected}. We also make the assumption that $f(z) + g(\overline z) = \gamma$, where $\gamma$ is some constant to be determined -- this turns out to be sufficient to solve~\eqref{eq:Corrected}.

We expand the metric around the Ricci-flat metric $G_{m\overline{n}} = \tilde{G}_{m\overline{n}} + \alpha'^3\delta G_{m\overline{n}}$, where the correction is $\mathcal{O}({\alpha'}^3)$. The correction to the Kähler potential $\delta K = (K - \tilde{K})/\alpha'^3$, to leading order in $\alpha'$, satisfies
\begin{equation}
    \alpha'^3\tilde{\triangle}\delta K = -\frac{(2\pi \alpha')^3}{4} \zeta(3) Q(\tilde G) - (2\pi)\gamma + \mathcal{O}(\alpha'^4). 
\end{equation}
We will drop the $\mathcal{O}(\alpha'^4)$ terms from now on, leaving them as understood.

As $X$ is compact, standard PDE theory requires that the source term must integrate to zero for a solution to exist. This leads to a condition on $\gamma$. Since $Q$ is the Euler density, we need to impose
\begin{equation}
    \gamma = -\frac{(2\pi \alpha')^3}{8\pi V} \zeta(3) \chi,
\end{equation}
where $V = \int_{X}(\det{\tilde{G}})$ is the volume of the Ricci-flat Calabi-Yau.

Putting the above together, the deviation from the Ricci-flat Kähler potential is, cancelling $\alpha'^3$ on both sides, given by the solution to
\begin{equation}\label{eqn:LapCorec}
    \tilde{\triangle}\delta K = \frac{(2\pi )^3}{4} \zeta(3)\left(\frac{\chi}{l_s^{6}\mathcal{V}} - Q(\tilde G)\right) \equiv \rho.
\end{equation}
Similar differential equations have been solved with machine learning methods in calculating physical Yukawa couplings~\cite{Constantin:2024yxh}. We follow the same approach by first taking the neural network that represents the Ricci-flat Kähler potential $\tilde{K}$ from the previous section, and represent the correction $\delta K$ with another neural network of the same type, trained with the loss functional\footnote{As a check, we also calculated the Euler character with the Ricci-flat metric using Q, and found it was correct to within $ <1 \%$ for several Calabi-Yau manifolds.}
\begin{equation}
    \mathcal{L}_{\alpha'}[\delta K] = \sum_{p\in X}|\tilde{\Delta} \delta K(p) - \rho(p)|.
\end{equation}

For both neural networks, we use architectures analogous to those introduced in~\cite{Douglas:2020hpv}. We find the trained neural network for $\delta K$, in the case of the Fermat quintic, solves equation~\ref{eqn:LapCorec} to within $\mathcal O (1\%)$ for a network with $10^4$ parameters.

\section{Computing the Change to the Massive Spectrum}\label{sec:MassShift}
\begin{figure}
    \centering
    \includegraphics[width=\linewidth]{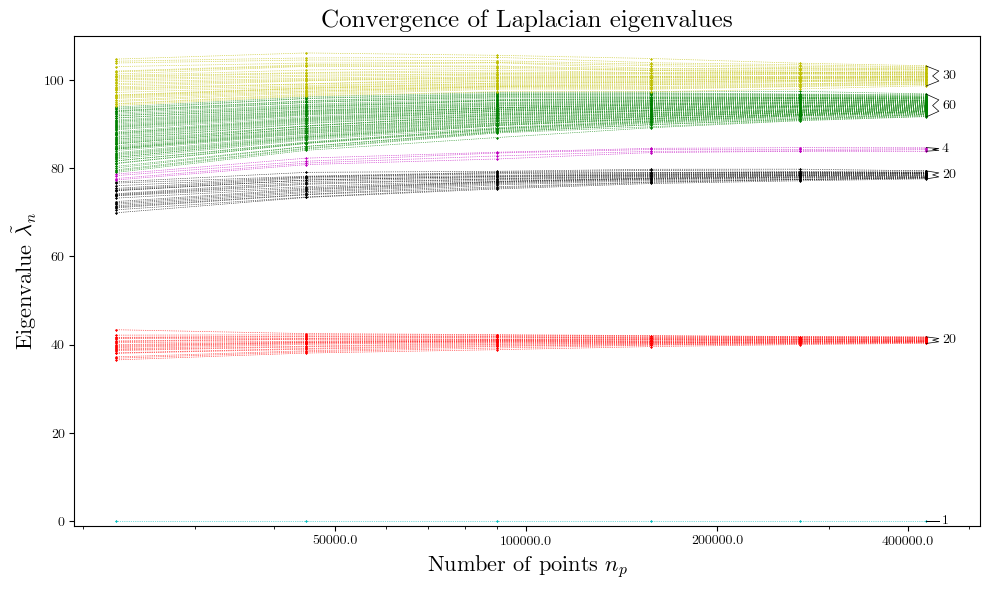}
    \caption{Spectrum of the scalar Laplacian on the Ricci-flat quintic, using a finite basis of functions constructed from the sections of $\mathcal O(3)$, as a function of the number of points used in the integration.  Note that the eigenvalues fall into representations of the discrete group of the Fermat quintic.}
    \label{fig:FlatSpectrum}
\end{figure}

We compute the spectrum of the scalar Laplacian in the same way as~\cite{Braun:2008jp}, where we take the sections $\{S_\alpha\}$ of $\mathcal{O}(3)$ on the quintic, and consider the partial basis of functions
\begin{equation}
    f_{\alpha,\bar{\beta}} = \frac{S_\alpha \bar{S}_{\bar\beta}}{(\sum_{p=0}^4 |X_p|^2)^3},
\end{equation}
where $X_p$ are the homogeneous coordinates of the quintic. Using the inner product formed with the Ricci-flat metric,
\begin{equation}\label{eq:innerProd}
    \braket{f|g} = \int_X \text{dVol}_{\tilde{G}}\bar{f}g,
\end{equation}
we can approximately compute the eigenvalues of the scalar Laplacian. These are indicated in figure~\ref{fig:FlatSpectrum}.

\begin{figure}
    \centering
    \includegraphics[width=\linewidth]{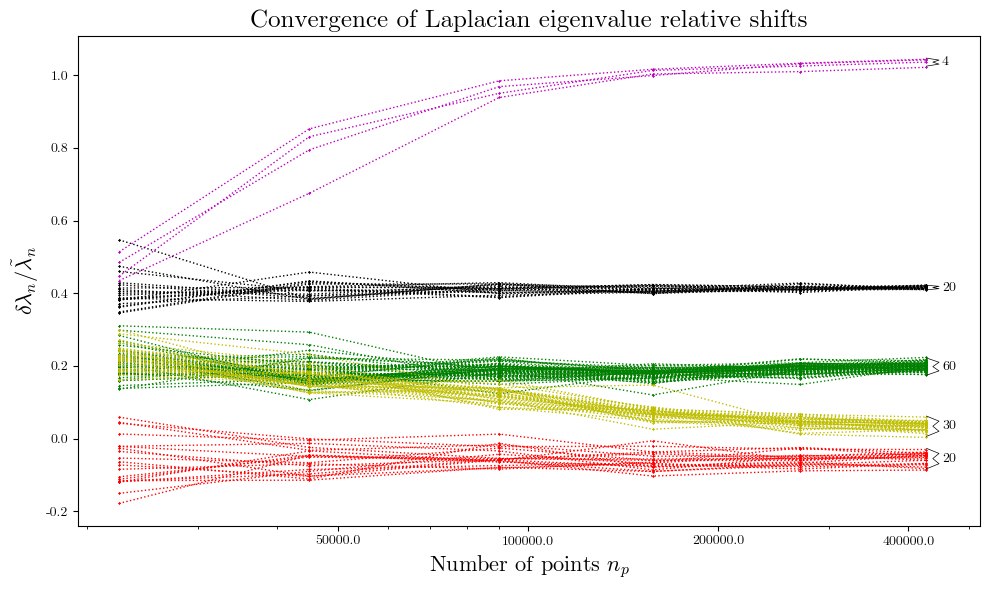}
    \caption{The relative shift to the spectrum of the scalar Laplacian on the Ricci-flat quintic, using a finite basis of functions constructed from the sections of $\mathcal O(3)$, as a function of the number of points used in the integration. The colours used in this figure match with those used in figure~\ref{fig:FlatSpectrum}.}
    \label{fig:SpectrumCorrec}
\end{figure}

The $\alpha'$ correction to the metric, leads to corrections to the Laplacian $\Delta$, eigenfunctions $\varphi_n$ and eigenvalues $\lambda_n$, which we call $\gamma$, $\omega$ and $\epsilon$ respectively:
\begin{align}
    \Delta &= \tilde{\Delta} + \alpha'^3\gamma,\\
    \lambda_n &= \frac{1}{\mathcal{V}^{\frac{1}{3}}} \tilde{\lambda}_n + \frac{\alpha'^3}{\mathcal{V}^{\frac{2}{3}}}\delta\lambda_n,\\
    \varphi_n &= \tilde{\varphi}_n + \alpha'^3\omega_n,
\end{align}
where the $\mathcal{V}$ factors are chosen such that $\tilde{\lambda}_n$ and $\delta\lambda_n$ are independent of $\mathcal V$. One finds that the change to the $n$-th eigenvalue, taking care that the inner product is also changed by $\alpha'$ corrections, is given by
\begin{equation}
    \delta\lambda_n =\frac{ \bra{\tilde\varphi_n}\gamma\ket{\tilde\varphi_n}}{\braket{\tilde\varphi_n|\tilde\varphi_n}}.
\end{equation}
We emphasise that this is with respect to the original, Ricci-flat, inner product~\eqref{eq:innerProd}. These corrections are shown in figure~\ref{fig:SpectrumCorrec} and table~\ref{tab:EvalShift}. It is interesting to note that there does not appear to be a correlation between the initial values of the eigenvalues, and the magnitude of the $\alpha'$ correction.

\begin{table}[h!]

\vspace{0.5cm}

\renewcommand{\arraystretch}{1.4} % Adjusts row spacing
\setlength{\tabcolsep}{5pt} % Adjusts column spacing
\begin{tabular}{|c||>{\centering\arraybackslash}p{0.7cm}|>{\centering\arraybackslash}p{0.7cm}|>{\centering\arraybackslash}p{0.7cm}|>{\centering\arraybackslash}p{0.7cm}|>{\centering\arraybackslash}p{0.7cm}|>{\centering\arraybackslash}p{0.7cm}|}
\hline
Rep. & \textbf{1} & \textbf{20} & $\textbf{20}_2$ & \textbf{4} & \textbf{60} & \textbf{30} \\ \hline\hline
\textbf{$\tilde{\lambda}_n$} & 0 & 41.0 & 78.5 & 84.3 & 94.2 & 100.9 \\ \hline
\textbf{$\delta\lambda_n$} & 0 & -2.2 & 32.6 & 87.3 & 18.7 & 3.5 \\ \hline
\end{tabular}
\caption{The eigenvalues of the Laplacian with respect to the Ricci-flat metric on the Fermat quintic, and the shift to these values from the $\alpha'$ correction. The eigenvalues form representations of the discrete symmetry group of the Fermat quintic~\cite{Braun:2008jp}.}
\label{tab:EvalShift}
\end{table}

\section{Conclusion}
In summary, this letter consisted of two parts, both of which concern numerical methods and $\alpha'$ corrections to string compactifications. The first was covered in section~\ref{sec:Valid}, where we highlight the use of numerical metrics in confirming control of the $\alpha'$ expansion for string compactifications: one can use numerical metrics to check conditions of the form given in equation~\ref{eqn:NewCond}. We claim these are stronger pointwise constraints than the usually considered $V \gg (2\pi)^6\alpha'^3$. In principle, these constraints should be checked in any string model, especially in the context of moduli stabilisation, where a specific point in moduli space has been chosen.

In the second part, we computed one of the many relevant corrections to numerical CY metrics: the change to the metric from the first $\alpha'$ correction, at tree-level in $g_{st}$, for IIB string theory in section~\ref{sec:MetricShift}. As a simple example of the use of this new metric, in section~\ref{sec:MassShift} we calculate the correction to the eigenvalues of the scalar Laplacian on the quintic, as indicated in figure~\ref{fig:SpectrumCorrec} and in table~\ref{tab:EvalShift}. 

This work initiates the use of taking machine learning methods for numerical metrics beyond the Ricci-flat metric, by considering one of the many corrections that need to be considered in a realistic string compactification. Looking forward, it is crucial to begin considering the backreaction from fluxes and local sources on these numerical metrics. These are important considerations both for realising the standard model within string theory, and for moduli stabilisation.
\vspace{0.2cm}
\section*{Acknowledgements}
The authors would like to thank James Halverson, Andre Lukas, Fernando Quevedo and Fabian Ruehle for helpful discussions. 
KFT would like to acknowledge the hospitality of IAIFI at the Massachusetts Institute of Technology, where a portion of this research took place. 

KFT is supported by the Gould-Watson Scholarship. TRH is supported by the National Science Foundation under Cooperative Agreement PHY-2019786 (The NSF AI Institute for Artificial Intelligence and Fundamental Interactions, \href{http://iaifi.org/}{http://iaifi.org/}). 

\bibliography{bibliography}
\bibliographystyle{inspire}

\end{document}